\documentstyle[prd,aps,epsf]{revtex}
\begin{document}
\draft
\preprint{}
\title{Interaction of massless Dirac field with a Poincar\'e gauge field}

\author{S. Nakariki and T. Masaki}
\address{Department of Applied Physics, Okayama University of Science, 
Ridai-cho 1-1, Okayama 700, Japan }

\author{K. Fukuma}
\address{Department of Control Engineering, Takuma National College of Technology, Takuma-cho Mitoyo-gun, Kagawa 769-11, Japan }

\author{T. Fukui}
\address{Mukogawa Women's University, Ikebiraki-cho, Nishinomiya, Hyougo 663, Japan}

\author{M. Mizouchi}
\address{Department of Liberal Arts and Science, Kurashiki University of Science and the Arts, Nishinoura tsurajima-cho, Kurashiki-shi, 712, Japan }

\author{T. Ohtani}
\address{Kansai Gaidai University, Hirakata 573, Japan}

\author{T. Tashiro}
\address{Department of Mechanical Engineering, Okayama University of Science, Ridai-cho 1-1, Okayama 700, Japan}

\date{\today}
\maketitle

\begin{abstract}
In this paper we consider a model of Poincar\'e gauge theory (PGT) in which a
 translational gauge field and a Lorentz gauge field are actually identified with the Einstein's
 gravitational field and a pair of ``Yang-Mills'' field and its partner, respectively.
In this model we re-derive some special solutions and take up one of
 them. The solution represents a ``Yang-Mills'' field without its
 partner field and the Reissner-Nordstr\"om type spacetime, which are
 generated by a PGT-gauge charge and its mass.
It is main purpose of this paper to investigate the interaction of massless Dirac fields with those fields.
As a result, we find an interesting fact that the left-handed massless Dirac fields
 behave in the different manner from the right-handed ones. This can be explained as to be caused by the direct interaction
of Dirac fields with the ``Yang-Mills'' field. Accordingly, the
 phenomenon can not happen in the behavior of the neutrino waves 
in ordinary Reissner-Nordstr\"om geometry.
The difference between left- and right-handed effects is calculated
 quantitatively, considering the scattering problems of the massless
 Dirac fields by our Reissner-Nordstr\"om type black-hole.

\end{abstract}


\section{Introduction}

Poincar\'e gauge theory was first founded by Utiyama\cite{utiyama}
and Kibble\cite{kibble}, and later developed by Hayashi\cite{hayashi},
Hehl and his collaborators\cite{hehl}. Since then, PGT has been studied
by many people. Most of them have adopted a model with nine  independent
parameters.  The physical meaning of those parameters has 
been examined until now only in the weak -field
approximation\cite{ohtani}. The good choice for parameters and also PGT
itself will, of course, have to be tested eventually through some
experiments. However, before doing so we need to know the behavior of strong Poincar\'e gauge fields and therefore to have some exact solutions of PGT with nine independent parameters. But it is very hard to solve exactly such the full-theory. Thus various models have been proposed and solved
by many authors\cite{example}. In those solutions there seems to exist
such ones that the property depends strongly on the structure of the gauge group and therefore have little dependence on the choice of 
parameters, i.e., on a peculiarity of the models. For example, in a few years ago\cite{nakariki} one of the
authors has obtained exact solutions (monopole solutions) in a model
where PGT can be actually identified with {\it complex} Einstein-Yang=Mills
theory.  The idea is based on an universal property which is an
extension of 't Hooft's original idea\cite{thooft}: any non-Abelian
theory can have a monopole solution if it has a compact covering group
in {\it its static limit}. Accordingly, the investigations based on the
solutions seem to give us some universal results. 

In this paper we shall re-derive some special solutions of above
ones, but in the different method from Ref\cite{nakariki}. 
For, owing to this method we can more naturally interpret two kinds of
integral (vector) constants, \(\vec{Q}_2,~\vec{Q}_1\) as the ``gauge
charges'' which creat the ``Yang-Mills'' field and its partner field,
respectively. And the magnitude \(|\vec{Q}_2|\) can be identified with
an electric charge, according to Kalb's anzatz\cite{kalb}. 
On the other hand, our gravitational field is generated by not only the
mass \(M\) but also two gauge charges  \(\vec{Q}_1,~\vec{Q}_2\) through
a combination \(q_{\ast}{}^2=-2(a_1+a_3)(\vec{Q}_1{}^2 - \vec{Q}_2{}^2)\).
And the spacetime structures can be classified by the signature
of \(q_{\ast}{}^2\) as follows: (I) if \(q_{\ast}{}^2=0\), then the spacetime is just Schwarzschildian
, (II) if \(q_{\ast}{}^2 >0\), Reissner-Nordstr\"omian, and (III) if
\(q_{\ast}{}^2 <0\), then Schwarzschildian-like.
The difference between (I) and (III) may, for example, be clarified by
considering the correction of the classical Kepler orbits. 
The details will be discussed in the forthcoming paper.

By the way,  it is well-known in the weak -field approximation that the
Poincar\'e gauge fields interact with the Dirac field through only the
axial part of the Lorentz gauge fields.  In our solutions the
Dirac fields are also able to interact with the gravitational field through
the gauge charge \(q_{\ast}{}^2\). In fact, in a region \(\Delta \equiv r^2
-2Mr+ q_{\ast}{}^2 >0\), our Dirac equation can be written as
\([\gamma^k\partial_k + \frac{i}4
\Delta_{mnk}\gamma^k\sigma^{mn}-\frac12 i
\frac{Q_{2[a]}}{\sqrt{\Delta}}\gamma^a \gamma_5 + im]\Psi =0\), where 
\(\gamma^k\) and \(\sigma^{mn}\) are Dirac matrices and their
combinations, respectively. And \(\Delta_{kmn}\) are Ricci's rotation
coefficients and \(Q_{2[a]}\) is an a-component of \(\vec{Q}_2\).
From this we can easily see that the Dirac fields can directly interact with 
``Yang-Mills'' field through the gauge charge \(\vec{Q}_2\), and not
directly interact with the partner field but only through the gravitational field.
This means that the direction of \(\vec{Q}_1\) has no effect on the
Dirac fields. And therefore, if we put \(\vec{Q}_1=0\) we shall be able to
consider the behavior of the Dirac field in the gravitating ``pure Yang-Mills''
field which is generated by the gauge charge \(\vec{Q}_2\) and the mass
\(M\). This is the motivation of this paper.  Incidentally, in this case the gravity is a type of the
Reissner-Nordstr\"om spacetime with the gauge charge \(\vec{Q}_2\) in
place of ordinary charges.

The investigation is done by using the Newman-Penrose formalism, in
terms of which the problems on the behavior of the neutrino waves in Kerr
geometry have been discussed mainly by
Chandrasekhar\cite{chandrasekhar}.
In this paper we are discussing our problems following to him as similarly as possible.  
As a result, we find an interesting fact that the left-handed
massless Dirac fields behave in a different manner from the right-handed 
ones. This can be considered as to be caused by the direct interaction
of Dirac fields with the ``pure Yang-Mills'' field. Accordingly, the
phenomenon can not happen in the behavior of the neutrino waves in ordinary
Reissner-Nordstr\"om geometry. 

In the next section some special solutions of PGT-monopoles are derived
again in a different way from Ref.\cite{nakariki}. In Sec. III we
consider the equations of massless Dirac fields interacting with 
our Poincar\'e gauge fields in terms of spinor forms. 
In Sec. IV we shall investigate the scattering problems of massless Dirac
fields by a Reissner-Nordstr\"om type black-hole 
with a gauge charge in place of normal charges. 
And the final section is devoted to conclusions.

\section{Some special solutions in PGT}

We start with the following Lagrangian for gauge fields\cite{hayashi}:\footnote{We here use the same notations as of Ref.\cite{nakariki}.}
\begin{eqnarray*}
 L_g &=& \alpha~{}^T{\cal C}_{kmn}{}^T{\cal C}^{kmn} + \beta~
{}^V{\cal C}_k {}^V{\cal C}^k + \gamma~{}^A{\cal C}_k {}^A{\cal C}^k  
+ a_1A_{kmnp}A^{kmnp} + a_2B_{kmnp}B^{kmnp} + \\
 &+& a_3C_{kmnp}C^{kmnp} + a_4E_{km}E^{km} + a_5G_{km}G^{km} + a_6F^2 + aF. \nonumber 
\end{eqnarray*}
Here ${}^T{\cal C}_{kmn},\cdots $ and $A_{kmnp},\cdots, F$ are the irreducible 
components of  translational and Lorentz gauge field strengths which are defined in terms of tetrad $b_k{}^\mu$ and Lorentz gauge fields $A_{km\mu}$ as 
\begin{eqnarray}
&& {\cal C}_{kmn}=b_n{}^\mu~{\cal C}_{km\mu}=2b_{k\mu,\nu}~b_{[m}{}^\mu~b_{n]}{}^\nu +2 A_{k[mn]},\\
&& F_{kmnp}=b_n{}^\mu b_p{}^\nu~F_{km\mu\nu}=2(A_{km\nu,\mu} + A_{kr\mu}~A^r{}_{m\nu})b_{[n}{}^\mu b_{p]}{}^\nu.
\end{eqnarray}
And $a, \alpha, \beta, \gamma$ and $a_i(i=1,2,\cdots,6)$ are ten constant parameters and it is well-known that only five of six $a_i$ are independent\cite{hayashi-shirafuji}. 

In this paper we make slightly more loose choice than in Ref.\cite{nakariki} as follows: $ \alpha + \frac23 a = \beta - \frac23 a = \gamma + \frac32 a =0$ and $ 4a_1=3a_2=2a_4=24a_6,~ a_5=2a_3 $. In this choice action integral can be rewritten as 
\begin{equation}
\label{Ig}
Ig=\int d^4x b L_g= \int d^4x b (aR + L_F)
\end{equation}
with 
\begin{equation}
L_F =a_1(A_{kmnp}A^{kmnp}+\frac43 B_{kmnp}B^{kmnp} + 
2 E_{km}E^{km} + \frac16 F^2) + a_3(C_{kmnp}C^{kmnp} + 2 G_{km}G^{km}), 
\end{equation}
where $R$ is a Riemann scalar curvature defined by the metric 
$g_{\mu\nu}=b_{k\mu}b^k{}_\nu$, and then $\frac1{2a}$ can be plausibly interpreted as Einstein's gravitational constant and $a_1$ and $a_3$ only are parameters to be determined by experiments. 

From this action (\ref{Ig}) we can derive the following {\it complex} Einstein-Yang-Mills equations (CEYM):
\begin{eqnarray}
&& {\cal G}^{\mu\nu} =\frac1{2a}T_{(L)}{}^{\mu\nu}, 
\label{einstein}\\ 
&& \vec{\cal F}^{\mu\nu}{}_{;\nu}-i \vec{\cal A}_\nu \times \vec{\cal F}^{\mu\nu}=0, \\
&& \vec{\cal F}^{\dag \mu\nu}{}_{;\nu} - i \vec{\cal A}_\nu \times \vec{\cal F}^{\dag \mu\nu}=0.
\end{eqnarray}
Here ${\cal G}^{\mu\nu}$ is the Einstein's tensor and $T_{(L)}{}^{\mu\nu}$ is an energy-momentum tensor for the Lorentz gauge field:
\begin{equation}
\label{energymomentum}
T_{(L)}{}^{\mu\nu}=b_k{}^\mu b_m{}^\nu ~T_{(L)}{}^{km}=b_k{}^\mu b_m{}^\nu ~(
-2~F_{pqn}{}^k~\frac{\partial L_F}{\partial F_{pqnm}} + \eta^{km}~L_F).
\end{equation}
And $\vec{\cal A}_\mu$ and $\vec{\cal F}_{\mu\nu}$ are a complex field 
and its strength, respectively,  which are made from $ A_{km\mu}$ and $ F_{km\mu\nu}$ as follows:  
\begin{equation}
 \vec{\cal A}_\mu = \vec{v}_\mu + i~ \vec{a}_\mu, \qquad \vec{\cal F}_{\mu\nu}=\vec{F}^P{}_{\mu\nu} + i~ \vec{F}^A{}_{\mu\nu},
\end{equation} 
where  
\begin{eqnarray*}
&& A_{0a\mu} \equiv v_{(a)\mu} \rightarrow  \vec{v}_\mu, \qquad \frac12\epsilon_{abc}A_{bc\mu} \equiv a_{(a)\mu}  \rightarrow  \vec{a}_\mu \\
&& F_{0a\mu\nu} \equiv F^P{}_{(a)\mu\nu} \rightarrow  \vec{F}^P{}_{\mu\nu} \qquad \frac12\epsilon_{abc}F_{bc\mu\nu} \equiv F^A{}_{(a)\mu\nu}  \rightarrow \vec{F}^A_{\mu\nu}.
\end{eqnarray*}
The symbol (;) means the ordinary covariant derivatives with Christoffel connection, and the quantities with $\dagger$ symbol mean the duals to the corresponding ones.

Now we put 
\begin{equation}
\label{veccalA}
\vec{\cal A}_\mu = \vec{\beta}{\cal A}_\mu
\end{equation}
where $\vec{\beta}$ is a constant vector in 3D complex space. Then we get 
\begin{equation}
 \vec{\cal F}_{\mu\nu} =\vec{\beta}~{\cal F}_{\mu\nu} \qquad \mbox{with} \quad {\cal F}_{\mu\nu}={\cal A}_{\nu,\mu} - {\cal A}_{\mu,\nu},
\end{equation}
and therefore we are led to the {\it complex} Einstein-Maxwell equations for the complex field ${\cal A}_\mu$:
\begin{mathletters}
\begin{eqnarray}
 && {\cal F}^{\mu\nu}{}_{;\nu} = 0,  \\
&& {\cal F}^{\dag \mu\nu}{}_{;\nu} = 0 . 
\end{eqnarray}
\end{mathletters}
From these equations it is very easy to find a solution similar to the ordinary Coulomb field. In fact, we can get from those equations the following exact solution with a complex vector gauge charge $\vec{\cal Q}_\ast (=\vec{Q}_1 + i \vec{Q}_2)$:
\begin{equation}
\label{cvcfield}
 \vec{\cal A}_{\mu} = \vec{\beta}{\cal A}_0 \delta^0{}_{\mu}= 
\frac{\vec{\cal Q}_\ast}r \delta^0{}_{\mu}, 
\end{equation} 
or 
\begin{equation}
\vec{a}_{\mu} = \frac{\vec{Q}_2}r \delta^0{}_{\mu }, \qquad \vec{v}_{\mu} 
= \frac{\vec{Q}_1}r \delta^0{}_{\mu}.
\end{equation}
The energy-momentum tensor of this field is calculated by (\ref{energymomentum})  to be 
\begin{equation}
 T_{(L)}{}^{km}=diag[\frac{q_\ast{}^2}{r^4},~\frac{q_\ast{}^2}{r^4},~-\frac{q_\ast{}^2}{r^4},~\frac{q_\ast{}^2}{r^4} ]
\end{equation}
with 
\begin{equation}
 q_{\ast}{}^2=- 2 (a_1 + a_3)(\vec{Q}_1{}^2 - \vec{Q}_2{}^2).
\end{equation}

Substituting above results to the Einstein equation (\ref{einstein}) 
and according to the similar procedure to the real one\cite{chandrasekhar}, 
we can easily get the following spacetime: 
\begin{equation}
\label{delta+}
 d s^2 =\frac{\Delta}{r^2}~(d x^0)^2 - \frac{r^2}{\Delta}~(d r)^2 - r^2~\{ (d \theta)^2 + (d \phi)^2~\sin^2\theta \} \quad \mbox{in a region of}~ \Delta > 0,
\end{equation}
or 
\begin{equation}
 d s^2 = \frac{4|\Delta|}{r^2}~du dv - r^2~\{ (d \theta)^2 + (d \phi)^2~\sin^2\theta \} \quad \mbox{in a region of}~ \Delta < 0,
\end{equation}
where $\Delta=r^2 - 2Mr + q_\ast{}^2 \stackrel{\rm def}{=}(r - r_{+})
(r - r_{-})$ with $r_{\pm}= M \pm \sqrt{M^2 - q_{\ast}{}^2}$.

As seen from the relation between $\Delta$ and $r$, above spacetime is classified by the value  of $q_\ast{}^2$: (I)~ If $q_\ast{}^2=0$, then we have just 
Schwarzschild spacetime. (II)~ If $0<q_\ast{}^2<M^2$, then the spacetime is like Reissner-Nordstr\"om (RN spacetime). (III)~ If $q_\ast{}^2<0$, then we get a spacetime having the similar structure to Schwarzschild. The difference between the cases (I) and (III) may, for example, be clarified by considering the correction of the classical Kepler orbits. In fact, under post-Newtonian approximation $\mu \ll 1,~\lambda \ll 1,~\kappa \ll 1$, we can find that the case (III) gives the advance in the perihelion per revolution\cite{nakariki2}, 
\begin{equation}
\delta \phi \sim 2\pi \{3\mu + \frac12\kappa +3(1+\frac14e^2)\lambda\},
\end{equation}
where $\mu=\frac{M}\ell,~\lambda=\frac{|q_\ast{}^2|}{\ell^2},~\kappa=\frac{|q_\ast{}^2|}{L^2}$ ,and $\ell,~L$ and $e$ are in turn the latus rectum, the angular momentum and the eccentricity of the orbit.

By the way, in the previous paper\cite{nakariki} we saw the fields 
${\vec a}_{\mu}, {\vec v}_{\mu}$ could be interpreted as "Yang-Mills" field 
and its partner field, respectively. Furthermore, according to Kalb\cite{kalb} the following quantity $F_{\mu\nu}$ was interpreted as the electromagnetic 
field tensor:
\begin{equation}
\vec{a}_{\nu,\mu} - \vec{a}_{\mu,\nu} + \vec{a}_{\mu} \times  \vec{a}_{\nu}
= F_{\mu\nu} \frac{\vec{a}_0}{|\vec{a}_0|}.
\end{equation}
Following this interpretation the magnitude of the gauge charge $|\vec{Q}_2|$ can be identified with an electric charge by which an electric field $E_k = F_{0k} = \frac{|\vec{Q}_2|}{r^2} \delta^1{}_{k}$ is created.

In this paper we take up the case (II) and consider the scattering
problems of the massless Dirac fields by the ``pure Yang-Mills'' fields,
when the Dirac field approaches to the outer event horizon $r_{+}$ of 
the R-N type black-hole far from the outside. To this purpose we choose 
as $\vec{Q}_1 = 0 ,  \vec{Q}_2=(0,0,Q_{2[3]})$ and $a_1 + a_3 > 0, $ and
also the tetrad $ b_k{}^\mu$ as 
\begin{eqnarray}
b_0{}^\mu &=& \left( \frac{3r^2 -2Mr + q_\ast{}^2}{2\sqrt{2} \Delta}, 
\frac{r^2 + 2M r - q_\ast{}^2}{2\sqrt{2} r^2}, 0, 0\right), \qquad 
b_1{}^\mu = \left(0, 0, \frac1r, 0\right), \nonumber \\
b_2{}^\mu &=& \left( 0, 0, 0, \frac1{r \sin\theta} \right), \qquad 
b_3{}^\mu = \left( \frac{r^2 +2Mr - q_\ast{}^2}{2\sqrt{2} \Delta}, 
\frac{3r^2 - 2M r + q_\ast{}^2}{2\sqrt{2} r^2}, 0, 0\right).  
\end{eqnarray}

\section{Dirac equation}

The gauge-invariant Lagrangian for the Dirac fields can be obtained from 
the original one by means of the replacement of the ordinary derivative 
$\Psi_{,k}$ by the covariant one $D_k \Psi=b_k{}^\mu (\Psi_{,\mu} + 
\frac{i}2 A_{mn\mu} S^{mn} \Psi)$. Thus action is given as 
\begin{equation}
\label{diracaction}
I_{Dirac} = \int d^4x b \left[\frac12 \bar{\Psi} \gamma^k D_k \Psi - \frac12 D_k \bar{\Psi} \gamma^k \Psi + i m \bar{\Psi}\Psi \right].
\end{equation}
Here $\gamma^k (k=0,1,2,3)$ are Dirac matrices and we shall adopt the representations:

\begin{equation}
\label{diracgamma}
\gamma_0=\left(\begin{array}{cc}
0 & 1 \\
1 & 0 
\end{array}
\right), \qquad \gamma_i =\left(\begin{array}{cc}
0 & - \sigma_i \\
\sigma_i & 0
\end{array}
\right), \qquad \gamma_5 = i \gamma^0 \gamma^1 \gamma^2 \gamma^3 
=\left(\begin{array}{cc}
1 & 0 \\
0 & - 1 
\end{array}
\right),
\end{equation}
where $ \sigma_i (i=1,2,3)$ are Pauli matrices.  Six generators $S^{km}$ are now given in terms of $\gamma^k$ as $S^{km} = -\frac12 \sigma^{km} = - \frac{i}4 [\gamma^k, \gamma^m]$.

Applying the variational principle to (\ref{diracaction}) we can derive the following equation:
\begin{equation}
\label{diraceq}
\left[\gamma^k (\nabla_k -\frac34 i {}^A{\cal C}_k \gamma_5) + i m\right]\Psi=0.\end{equation}
Here ${}^A{\cal C}_k$ is an axial vector part of the translational gauge field strength: ${}^A{\cal C}_k = \frac1{3!}\epsilon_{kmnp}{\cal C}^{mnp}$. It should be also remarked that a new covariant derivative $\nabla_k \Psi$ is introduced here. This is defined in terms of Ricci's rotation coefficients $\Delta_{km\mu}$ instead of $A_{km\mu}$ in $D_k \Psi$ as $ \nabla_k \Psi = b_k{}^\mu (\partial_\mu \Psi + \frac{i}4 \Delta_{mn\mu} \sigma^{mn} \Psi)$. 
 
It is convenient for our purpose to resolve above the 4-components Dirac 
equation (\ref{diraceq}]) to two 2-components equations. Using the well-known 
technique\cite{spinortech} these are given as  
\begin{mathletters}
\label{twocompoeq}
\begin{eqnarray}
&& \partial_{\dot{A}B}(\psi_L)^B + \Gamma^B{}_{C\dot{A}B} (\psi_L)^C -\frac34 i {}^A{\cal C}_{\dot{A}B} (\psi_L)^B + \frac{i m}{\sqrt{2}}(\bar{\psi}_R)_{\dot{A}} = 0, \\
&& \partial_{\dot{A}B}(\psi_R)^B + \Gamma^B{}_{C\dot{A}B} (\psi_R)^C -\frac34 i {}^A{\cal C}_{\dot{A}B} (\psi_R)^B + \frac{i m}{\sqrt{2}}(\bar{\psi}_L)_{\dot{A}} = 0.\end{eqnarray}
\end{mathletters}
Here we put 
\begin{eqnarray}
&& \Psi=\left(\begin{array}{c}
\psi_L \\
\psi_R 
\end{array}
\right),\qquad \partial_{\dot{A}B} = b_k{}^\mu 
\sigma^k{}_{\dot{A}B}\partial_\mu, \qquad {}^A{\cal C}_{\dot{A}B}= b_k{}^\mu \sigma^k{}_{\dot{A}B}{}^A{\cal C}_\mu, \\
&& \Gamma_{AB\dot{C}D}= - \frac12 \Delta_{km\mu}b_n{}^\mu
\sigma^{k\dot{E}}{}_A \sigma^m{}_{\dot{E}B} \sigma^n{}_{\dot{C}D},
\end{eqnarray}
where $\sigma^k{}_{\dot{A}B} (k=0,1,2,3)$ are multiples of identity and 
Pauli matrices by $\frac1{\sqrt{2}}$. 

Substituting the results of previous section for above equations (\ref{twocompoeq}) and putting the dependency on $t$ and $\phi$ as $e^{i({\hat \sigma} t + \hat{m} \phi)}$ we can get a set of equations
\begin{mathletters}
\label{diracequations}
\begin{eqnarray}
&& {\cal D}_0 f_1 + \frac1{\sqrt{2}} {\cal L}_{\frac12} f_2 =0,\label{eq1} \\
&& \Delta ({\cal D}^{\dag}{}_{\frac12} - i \frac{r}{\Delta} Q_{2[3]}) f_2 - \sqrt{2} {\cal L}^{\dag}{}_{\frac12} f_1 = 0, \label{eq2} \\
&& {\cal D}_0 g_2 - \frac1{\sqrt{2}} {\cal L}^{\dag}{}_{\frac12} g_1 =0,\label{eq3} \\
&& \Delta ({\cal D}^{\dag}{}_{\frac12} + i \frac{r}{\Delta} Q_{2[3]}) g_1 + \sqrt{2} {\cal L}_{\frac12} g_2 = 0. \label{eq4} 
\end{eqnarray}
\end{mathletters}
Here we define 
\begin{mathletters}
\begin{eqnarray}
&& f_1 = r (\psi_L)^0, \quad f_2 = (\psi_L)^1, \quad g_1=
(\bar{\psi}_R)^{\dot{1}}, \quad g_2 = - r (\bar{\psi}_R)^{\dot{0}}, \\
&& {\cal D}_n = \partial_r + i\frac{\hat{\sigma} r^2}{\Delta} + 2n\frac{r-M}{\Delta},\qquad {\cal D}^{\dag}{}_n = \partial_r - i\frac{\hat{\sigma} r^2}{\Delta} + 2n\frac{r-M}{\Delta}, \\
&& {\cal L}_n = \partial_{\theta} + n\cot\theta + \hat{m}\csc\theta, \qquad 
{\cal L}^{\dag}{}_n = \partial_{\theta} + n\cot\theta - \hat{m}\csc\theta.
\end{eqnarray}
\end{mathletters}
In order to solve the equations (\ref{diracequations}) we put 
\begin{eqnarray}
&& f_1=R_{-\frac12}(r) S_{-\frac12}(\theta), \qquad f_2=R_{+\frac12}(r) S_{+\frac12}(\theta), \nonumber \\
&& g_1=\tilde{R}_{+\frac12}(r) \tilde{S}_{-\frac12}(\theta), \qquad g_2=\tilde{R}_{-\frac12}(r) \tilde{S}_{+\frac12}(\theta).
\end{eqnarray}
Then the equations (\ref{eq1}) and (\ref{eq2}) are separated to produce 
\begin{mathletters}
\label{equation1}
\begin{eqnarray}
&& {\cal D}_0 R_{-\frac12} = \lambda_1 R_{+\frac12}, \label{eqa} \\
&& \Delta [{\cal D}^{\dag}{}_{\frac12} - i \frac{r}{\Delta} Q_{2[3]}] R_{+\frac12} = \lambda_2 R_{-\frac12}, \label{eqb} \\
&& {\cal L}_{\frac12} S_{+\frac12} = - \sqrt{2} \lambda_1 S_{-\frac12}, \label{eqc} \\
&& {\cal L}^{\dag}{}_{\frac12} S_{-\frac12} =  \frac1{\sqrt{2}} \lambda_2 S_{+\frac12}, \label{eqd} 
\end{eqnarray}
\end{mathletters}
where $\lambda_1$ and $\lambda_2$ are separation constants. Similarly 
the following equations can be derived from (\ref{eq3}) and (\ref{eq4}):
\begin{mathletters}
\label{equation2}
\begin{eqnarray}
&& {\cal D}_0 \tilde{R}_{-\frac12} = \lambda_3 \tilde{R}_{+\frac12}, \label{eqe} \\
&& \Delta [{\cal D}^{\dag}{}_{\frac12} + i \frac{r}{\Delta} Q_{2[3]}] \tilde{R}_{+\frac12} = \lambda_4 \tilde{R}_{-\frac12}, \label{eqf} \\
&& {\cal L}_{\frac12} \tilde{S}_{+\frac12} = - \frac1{\sqrt{2}} \lambda_4 \tilde{S}_{-\frac12}, \label{eqh} \\
&& {\cal L}^{\dag}{}_{\frac12} \tilde{S}_{-\frac12} =\sqrt{2} \lambda_3 \tilde{S}_{+\frac12}. \label{eqg} 
\end{eqnarray}
\end{mathletters}

We first consider the angular parts of the equations (\ref{equation1}) and (\ref{equation2}). Operating ${\cal L}^{\dag}{}_{\frac12}$ and ${\cal L}_{\frac12}$ on (\ref{eqc}) and (\ref{eqd}), respectively,  from the left-hand side we get in turn 
\begin{mathletters}
\begin{eqnarray}
&& {\cal L}^{\dag}{}_{\frac12}{\cal L}_{\frac12} S_{+\frac12}(\theta)= -\lambda_1 \lambda_2 S_{+\frac12}(\theta), \label{angleeq1} \\
&& {\cal L}_{\frac12}{\cal L}^{\dag}{}_{\frac12} S_{-\frac12}(\theta)= -\lambda_1 \lambda_2 S_{-\frac12}(\theta). \label{angleeq2}
\end{eqnarray}
\end{mathletters}
In the same way we shall get the following equations from (\ref{eqh}) and (\ref{eqg}) :  
\begin{mathletters}
\begin{eqnarray}
&& {\cal L}^{\dag}{}_{\frac12}{\cal L}_{\frac12} \tilde{S}_{+\frac12}(\theta)= -\lambda_3 \lambda_4 \tilde{S}_{+\frac12}(\theta), \label{angleeq3} \\
&& {\cal L}_{\frac12}{\cal L}^{\dag}{}_{\frac12} \tilde{S}_{-\frac12}(\theta)= -\lambda_3 \lambda_4 \tilde{S}_{-\frac12}(\theta). \label{angleeq4}
\end{eqnarray}
\end{mathletters}

Replacing $\theta$ by $\pi - \theta$ in (\ref{angleeq1}) and (\ref{angleeq3}) and using a relation ${\cal L}^{\dag}{}_n(\theta) = - {\cal L}_n(\pi-\theta)$, and comparing the results with (\ref{angleeq2}) and (\ref{angleeq4}) we 
can conclude 
\begin{equation}
 S_{+\frac12}(\pi - \theta) = S_{-\frac12}(\theta), \qquad 
 \tilde{S}_{+\frac12}(\pi - \theta) = \tilde{S}_{-\frac12}(\theta).
\end{equation}
Furthermore, using these relations in (\ref{eqc}) and (\ref{eqg}) and 
comparing the results with (\ref{eqd}) and (\ref{eqh}) we get the relations 
\begin{equation}
\lambda_2 = 2 \lambda_1 , \qquad 
\lambda_4 = 2 \lambda_3 .
\end{equation}
Then the comparison of  (\ref{angleeq2}) and (\ref{angleeq4}) with an 
equation satisfied by spin(-1/2)-weighted spherical harmonics: 
${\cal L}_{\frac12}{\cal L}^{\dag}{}_{\frac12}{}_{-\frac12}Y_{\ell m}= 
- (\ell + \frac12)^2 {}_{-\frac12}Y_{\ell m}$ gives us the results 
$ S_{-\frac12}= {}_{-\frac12}Y_{\ell m}$ with $ \lambda_1=\frac12 \lambda_2=\frac1{\sqrt{2}}(\ell + \frac12)$ and $ \tilde{S}_{-\frac12}= {}_{-\frac12}Y_{\ell m}$ with $ \lambda_3=\frac12 \lambda_4=\frac1{\sqrt{2}}(\ell + \frac12)$. 

We are now in a position to consider the radial parts of equations 
(\ref{equation1}) and (\ref{equation2}). We first notice the relations 
$\Delta^{\frac12} {\cal D}^{\dag}{}_{\frac12} = {\cal D}^{\dag}{}_0 \Delta^{\frac12}$ and $\Delta {\cal D}_1 = {\cal D}_0 \Delta$. Applying the former to (\ref{eqb}) and putting as 
\begin{equation}
R_{\pm\frac12}=\frac{2^{\frac14 \pm \frac14}}2(Z_{+} \pm Z_{-})
\Delta^{-(\frac14 \pm \frac14)}\exp\left[\frac12 i \int 
\frac{r}{\Delta}Q_{2[3]} dr \right]
\end{equation}
we can derive the following equations from (\ref{eqa}) and (\ref{eqb}):
\begin{equation}
\label{radialeq1}
\left( \frac{d}{d r_{\ast}} \mp W \right) Z_{\pm} = i \hat{\sigma} Z_{\mp}.
\end{equation}
Here we have introduced a new variable $r_{\ast}$ and a new quantity $W$, which are 
defined by 
\begin{equation}
r_{\ast} = r + \frac{r_{+}(r_{+} + \frac{Q_{2[3]}}{2 \hat{\sigma}})}
{r_{+} - r_{-}}\ln |r - r_{+}| - \frac{r_{-}(r_{-} + \frac{Q_{2[3]}}
{2 \hat{\sigma}})}{r_{+} - r_{-}}\ln |r - r_{-}| 
\end{equation}
and
\begin{equation}
W = \frac{(\ell + \frac12) \frac{\Delta^{\frac12}}{r^2}}{1 + \frac{Q_{2[3]}}{2 \hat{\sigma} r}}.
\end{equation}
In the same way we can also get the following equations from (\ref{eqe}) and 
(\ref{eqf}):
\begin{equation}
\label{radialeq2}
\left( \frac{d}{d \tilde{r}_{\ast}} \mp \tilde{W} \right) \tilde{Z}_{\pm} 
=- i \hat{\sigma} \tilde{Z}_{\mp}.
\end{equation}
Here we have also introduced new quantities $\tilde{Z}_{\pm}$  which are defined through  
\begin{equation}
\tilde{R}_{\pm\frac12}=\frac{2^{\frac14 \pm \frac14}}2(\tilde{Z}_{+} \pm 
\tilde{Z}_{-}) \Delta^{-(\frac14 \pm \frac14)}\exp\left[ \frac12 i \int 
\frac{r}{\Delta}Q_{2[3]} dr \right],
\end{equation}
\begin{equation}
\tilde{r}_{\ast} = r + \frac{r_{+}(r_{+} - \frac{Q_{2[3]}}{2 \hat{\sigma}})}
{r_{+} - r_{-}}\ln |r - r_{+}| - \frac{r_{-}(r_{-} - \frac{Q_{2[3]}}
{2 \hat{\sigma}})}{r_{+} - r_{-}}\ln |r - r_{-}| 
\end{equation}
and  
\begin{equation}
\tilde{W} = \frac{(\ell + \frac12) \frac{\Delta^{\frac12}}{r^2}}
{1 - \frac{Q_{2[3]}}{2 \hat{\sigma} r}}.
\end{equation}

For later discussion we would like to note here that the gauge charge 
$Q_{2[3]}$ gives a different contribution to the left-handed massless 
Dirac fields from the right-handed ones.

\section{Scattering of Dirac fields by a black-hole with a PGT-gauge charge}

As stated in final part of section II, we consider here a scattering
problem of the massless Dirac fields by a Reissner-Nordstr\"om type
black-hole having a gauge charge $Q_{2[3]}$ and a mass {\it M}.
However, we shall constrain ourselves to discuss only such a case that
the massless Dirac fields approach to the event horizon $r_{+}$ of the
black-hole far from the outside. 

To this purpose we first note a fact that the following two Schr\"odinger equations 
can be made from (\ref{radialeq1}) and (\ref{radialeq2}):
\begin{eqnarray}
\label{schroedinger1}
&& \left(- \frac{d^2}{d r_{\ast}{}^2} + V_{\pm} \right) Z_{\pm} = \hat{\sigma}^2 Z_{\pm} \qquad \mbox{with} \quad V_{\pm}= W^2 \pm \frac{d W}{d r_{\ast}}, \\
\label{schroedinger2}
&& \left(- \frac{d^2}{d \tilde{r}_{\ast}{}^2} + \tilde{V}_{\pm} \right) 
\tilde{Z}_{\pm} = \hat{\sigma}^2 \tilde{Z}_{\pm} 
\qquad \mbox{with} \quad \tilde{V}_{\pm}= \tilde{W}^2 \pm \frac{d \tilde{W}}{d \tilde{r}_{\ast}}.
\end{eqnarray}

From these equations we can get at once the following conserved Wronskians:
\begin{equation}
\frac{d [Z_{\pm}, Z^{\ast}{}_{\pm}]}{d r_{\ast}} =0, \qquad \frac{d [\tilde{Z}_{\pm}, \tilde{Z}^{\ast}{}_{\pm}]}{d \tilde{r}_{\ast}}=0,
\end{equation}
where the Wronskians are defined as 
\begin{equation}
[Z_{\pm}, Z^{\ast}{}_{\pm}]=Z_{\pm} \frac{dZ_{\pm}{}^{\ast}}{dr_{\ast}}  - Z_{\pm}{}^{\ast} \frac{dZ_{\pm}}{dr_{\ast}}, \qquad   [\tilde{Z}_{\pm}, \tilde{Z}^{\ast}{}_{\pm}]= \tilde{Z}_{\pm}  \frac{d\tilde{Z}_{\pm}{}^{\ast}}{d\tilde{r}_{\ast}} - \tilde{Z}_{\pm}{}^{\ast} \frac{d\tilde{Z}_{\pm}}{d\tilde{r}_{\ast}}.
\end{equation}
And also the following relations are obtained using (\ref{radialeq1}) and (\ref{radialeq2}): 
\begin{equation}
Z_{+}{}^{\ast} Z_{-} + Z_{-}{}^{\ast} Z_{+} =\frac1{-i \hat{\sigma}} [Z_{+}, Z^{\ast}{}_{+}],
\qquad 
\tilde{Z}_{+}{}^{\ast}\tilde{Z}_{-} + \tilde{Z}_{-}{}^{\ast} \tilde{Z}_{+} =\frac1{-i \hat{\sigma}}[\tilde{Z}_{+}, \tilde{Z}^{\ast}{}_{+}].
\end{equation}

Before going forward, we must discuss about the conserved current of massless Dirac field. First, it should be remarked that the Dirac action (\ref{diracaction}) is invariant under the phase transformation $\Psi'=e^{i \alpha} \Psi \quad (\alpha =\mbox{real constant})$, so that we can get a continuity equation 
\begin{equation}
\label{continuityeq}
\partial_{\mu}(b J^{\mu})=0,
\end{equation}
where $J^{\mu}=b_k{}^{\mu}\bar{\Psi} \gamma^k \Psi$.  
$J^{\mu}$ can also be written, using the representation (\ref{diracgamma}) for Dirac gamma matrices, as 
\begin{equation}
J^{\mu} = J_L{}^{\mu} + J_R{}^{\mu}, 
\end{equation}
where $J_L{}^{\mu} = b_k{}^{\mu} \psi_L{}^{\dagger} \sigma^k \psi_L,\quad J_R{}^{\mu}=
b_k{}^{\mu} \psi_R{}^{\dagger} \bar{\sigma}^k \psi_R$ with $\sigma^k=\{I, \sigma_i(i=1,2,3)\}$ 
and $\bar{\sigma}^k=\{I, -\sigma_i (i=1,2,3)\}$. 
Thus the conserved net current of massless Dirac particles is calculated on account of the spherical symmetry to be 
\begin{equation}
\frac{\partial N}{\partial t}=- \int_0^{2\pi}\int_0^{\pi} (J_L{}^r + J_R{}^r )b d\theta d\phi =\frac{i\pi}{\sigma}\{
[Z_{+}, Z^{\ast}{}_{+}] + [\tilde{Z}_{+}, \tilde{Z}^{\ast}{}_{+}] \}, 
\end{equation}
where the first term of the right-hand side represents the contribution
of the left-handed Dirac fields and the second of the right-handed ones.

Hereafter, we assume $Q_{2[3]} <0$. This assumption can be done without the lose of generality, because the change of sign means only the 
exchange of a role of the left- and right- handed fields.
To go ahead, we further need to distinguish the following two cases:  (1)  $\hat{\sigma}>\sigma_s$ and (2) $0<\hat{\sigma}<\sigma_s$, where $\sigma_s$ is defined as $\sigma_s=-\frac{Q_{2[3]}}{2r_{+}}$.  In (1)  both $r_{\ast}$ and $\tilde{r}_{\ast}$ are single-valued functions over all range of $ r_{+} < r < \infty$ and both potentials $V_{\pm}$ and $\tilde{V}_{\pm}$ are continuous and short-range. On the other hand, in (2)  $\tilde{r}_{\ast}$ is also a single-valued function and the potential $\tilde{V}_{\pm}$ is continuous over all range of $ r_{+} < r < \infty$. However, $r_{\ast}$ has a minimum $ r_{\ast min}$ at $ r=\frac{\sigma_s}{\hat{\sigma}}r_{+}$. Therefore $ r_{\ast}$ is a double-valued function over the range $ r_{+} < r < \infty$ and the potentials $V_{\pm}$  become singular at  $ r_{\ast min}$. 
From this fact  we may expect the existence of the socalled super-radiance which is well-known in a scattering problem of an electromagnetic field by a Kerr black-hole. However, this being not so can be easily shown in exactly the same way as a case of scattering of massless Dirac fields by a Kerr black-hole\cite{chandrasekhar}.  

\begin{figure}
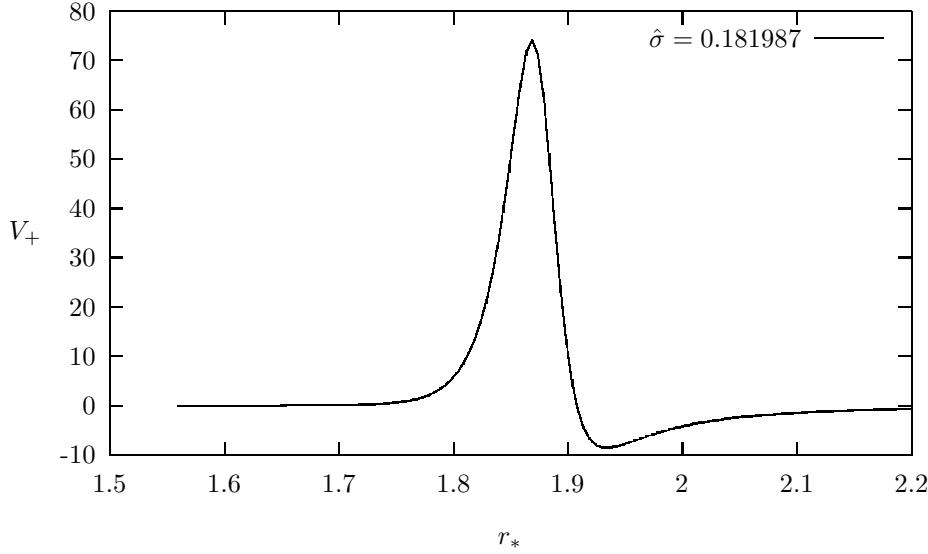

\setlength{\unitlength}{0.240900pt}
\ifx\plotpoint\undefined\newsavebox{\plotpoint}\fi
\sbox{\plotpoint}{\rule[-0.200pt]{0.400pt}{0.400pt}}%


\caption{A potential barrier, surrounding the RN type black-hole $(M=1)$ with a gauge charge $q_{\ast} =0.5$ and $\sigma_s=-\frac{Q_{2[3]}}{2r_+}=0.18099$ for the incidence of massless left-handed Dirac field with $\ell =0.5$ and the frequency $\hat{\sigma}=0.181987$.}
\label{fig1}
\end{figure}

\begin{figure}
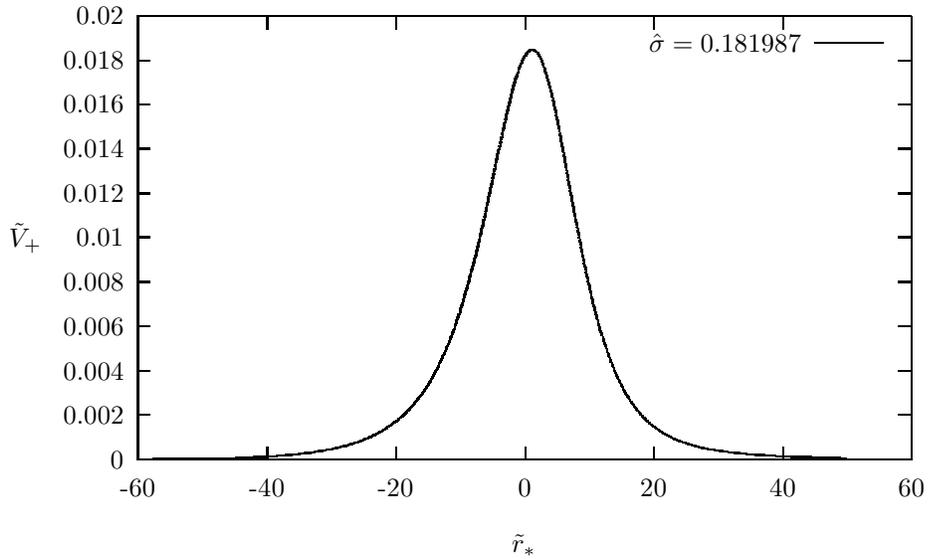

\setlength{\unitlength}{0.240900pt}
\ifx\plotpoint\undefined\newsavebox{\plotpoint}\fi
\sbox{\plotpoint}{\rule[-0.200pt]{0.400pt}{0.400pt}}%
%

\caption{A potential barrier for the right-handed Dirac field under the same conditions as the left-handed one.}
\label{fig2}
\end{figure}

In this paper we pick up only the case (1) and concretely calculate the reflection and transmission coeffcients for both left- and right-handed incident massless Dirac fields. 
In doing so we first remark that the wave functions $Z_{\pm} $ (or
$\tilde{Z}_{\pm} $ ) give the same reflection and transmission
coefficients because of the relations (\ref{radialeq1}) (or (\ref{radialeq2})). Therefore it is enough for us to consider only the scatterings by the potentials $ V_{+}$ and $ \tilde{V}_{+}$. 
These potentials tend to decrease whenever the frequency of incident fields $ \hat{\sigma}$ is increased.  (A prototype of these potentials is drawn in Fig.\ref{fig1} and \ref{fig2}.)  Accordingly, we can see, before the calculation, that the reflection coefficients decrease (and the transmission coefficients increase), in accordance with the increase of the frequencies of incident fields. 

The calculations were performed by using the 6th order Runge-Kutta
method and putting as $ \ell=\frac12,~M=1, ~Q_{2[3]}=-0.675470,
\hbar=1,~c=1, ~q_{\ast}=0.5$ and Einstein's gravitational constant being 
unity. The results are summarised in a table \ref{table1}.

\section{Summary and conclusions}

In this paper we have adopted a model of Poincar\'e gauge theory
which can be actually identified with {\it complex} Einstein-Yang=Mills
theory. And we have derived again a solution which can be interpreted as
the ``Yang-Mills'' fields generated by the gauge charge \(\vec{Q}_2\)
and the spacetime curved by Reissner-Nordstr\"om type black-hole. 
However, the solution is special one of more general set of
solutions\cite{nakariki} which are derived from the universal idea 
following to 'tHooft\cite{thooft}. 
Accordingly, we can expect that the investigations will give us the qualitative but universal results. 

Then we have investigated the behavior of the massless Dirac fields in
the external Poincar\'e gauge field mentioned above, namely 
the scattering problems of massless Dirac fields by the
gravitating ``Yang-Mills'' fields generated by the gauge charge 
 \(\vec{Q}_2\) and the mass \(M\).

 The investigation has been done by using the standard spinor techniques and following to
Chandrasekhar's method\cite{chandrasekhar}. As a result, we have found an interesting 
fact that the left-handed massless Dirac fields behave in a different
manner from the right-handed ones, owing to the gauge charge $Q_{2[3]}$.
The fact has been also examined concretely by setting in turn for
$Q_{2[3]},~q_{\ast}$ and $M$ to $-0.675470,~ 0.5$ and $1$. The results are summarised in the
table \ref{table1}, showing that more right-handed massless Dirac fields can
come near the outer event-horizon of the Reissner-Nordstr\"om type
black-hole than the left-handed ones if $Q_{2[3]}<0$.

Although in this paper we have restricted ourselves to a massless mode
of Dirac fields, it is an open question whether the same procedures
can be applied to the massive modes or not. This problem should be
considered before too long.

\newpage
\begin{table}
\caption{Reflection coefficients for $ \ell =0.5$ massless
 left- and right-handed Dirac particles incident on a Reissner-Nordstr\"om black-hole with a gauge charge $q_{\ast} =0.5,~ \sigma_s=-\frac{Q_{2[3]}}{2r_+}=0.18099$ and a mass $M=1$.}
\label{table1}
\begin{tabular}{cccccc}
$\hat{\sigma}$ & $R_{left}$ & $R_{right}$ & $\hat{\sigma}$ & $R_{left}$
 & $R_{right}$ \\
\tableline
 0.181987  & 0.97524 & 0.033468 & 0.220987 & 0.94419 & 0.0094065 \\ 
 0.182987 & 0.97477 & 0.032379  & 0.230987 & 0.92996 & 0.0068388 \\   
 0.183987 & 0.97427 & 0.031338  & 0.240987 & 0.91178 & 0.0049839 \\
 0.184987 & 0.97377 & 0.030319  & 0.250000 & 0.89127 & 0.0037564 \\
 0.185987 & 0.97326 & 0.029334  & 0.250987 & 0.88874 & 0.0036428  \\
 0.186987 & 0.97274 & 0.028392  & 0.260987 & 0.85994 & 0.0026686 \\
 0.187987 & 0.97220 & 0.027471  & 0.300000 & 0.67973 & 0.0008104 \\
 0.188987 & 0.97165 & 0.026580  & 0.350000 & 0.34543 & 0.0001834 \\
 0.189987 & 0.97109 & 0.025719  & 0.400000 & 0.11622 & 0.0000430 \\
 0.200987 & 0.96384 & 0.017949  & 0.450000 & 0.03212 & 0.0000104 \\
 0.210987 & 0.95525 & 0.012978  &          &         &  \\
\end{tabular}

\end{table}

\end{document}